\documentclass{article}
\parskip=\medskipamount

\title{Construction of Lagrangian local symmetries for general quadratic theory}
\author{A.A. Deriglazov\footnote{alexei@ice.ufjf.br ~ On leave of
absence from Dept. Math. Phys., Tomsk Polytechnical University,
Tomsk, Russia.}}
\date{Dept. de Matematica, ICE, Universidade Federal de Juiz de Fora,\\
MG, Brasil.}
\begin{document}
\maketitle
\large

\begin{abstract}
We propose a procedure which allows one to construct local symmetry generators of general quadratic  
Lagrangian theory. Manifest recurrence relations for 
generators in terms of so-called structure matrices of the Dirac formalism are obtained. The procedure 
fulfilled in terms of initial variables of the theory, and do not implies either separation of constraints 
on first and second class subsets or any other choice of basis for constraints.
\end{abstract}

\noindent
%{\bf PAC codes:} \\
%{\bf Keywords:} Hamiltonian systems with constraints, Gauge theories

\section{Introduction}

Relativistic theories are usually formulated in manifestly covariant form, i.e. in the form with linearly 
realized Lorentz group. It is achieved by using of some auxiliary variables, which implies appearance of 
local (gauge) symmetries in the corresponding Lagrangian action. Investigation of the symmetries is 
essential part of analysis of both classical and quantum versions of a theory. 
Starting from pioneer works on canonical quantization of singular theories [1-3], 
one of the most intriguing problems is search for constructive procedure 
which allows one to find the local symmetries from known Lagrangian or Hamiltonian 
formulation [4-10]. For a theory with first class constraints only, the problem has been discussed 
in [4, 5]. Symmetry structure (classification and proof on existence of irreducible complete set of gauge 
generators) for a general singular theory has been obtained in [6-8]. In particular, it was shown how 
one can find irreducible complete set of Hamiltonian gauge generators for general quadratic theory [7], 
as well as for general singular theory [8]. 

In the present work we propose an alternative to [7] procedure to construct Lagrangian local symmetries 
for the case of general quadratic theory\footnote{More exactly, our consideration is restricted to a theory 
with rank of the matrix $\{\Phi_1, \Phi_p\}$ be constant in vicinity of phase space point under  
consideration ($\Phi_p$ is a set of $p$-stage constraints).}. 
Total number of independent symmetries (see Sect. 5), which can be find by using of our procedure,   
coincides with number of Lagrangian multipliers remaining undetermined in the Dirac procedure (completeness 
of the set will not be discussed here). Our method is based on analysis of Noether identities in the 
Hamiltonian form, the latter has been obtained in our works [9, 10]. 
Some characteristic properties of our procedure are:
1) The procedure do not requires separation of Hamiltonian constraints on first and second class 
subsets, which is may be the most surprising result of the work.
2) The procedure do not requires choice of some special basis for constraints. 
3) All the analysis is fulfilled in terms of initial variables. 

To describe final result of the work, let us fix some notations. 
We consider singular Lagrangian theory  
$(A=1,2,\cdots [A])$
\begin{eqnarray}\label{201}
S=\int d\tau L(q^A,\dot q^A), \qquad 
rank\frac{\partial^2 L}{\partial\dot q^A\partial\dot q^B}= [i]<[A]. 
\end{eqnarray}
According to Dirac [1], Hamiltonian formulation of the theory is obtained as follow. 
First stage of Hamiltonization procedure is to define equations for   
the momenta $p_A$: $ ~ p_A=\frac{\partial L}{\partial \dot q^A} ~$. Being     
considered as algebraic equations for determining 
of velocities $\dot q^A$, $[i]$ equations 
can be resolved for $\dot q^i$ and then substituted into the remaining ones. By construction, the 
resulting equations do not depend on $\dot q^A$ and are called primary constraints 
$\Phi_\alpha(q, p), ~ \alpha=1, 2, \ldots , [\alpha]$ 
of the Hamiltonian formulation. The equations $p_A=\frac{\partial\bar L}{\partial \dot q^A}$ 
are then equivalent to the system   
$\dot q^i=v^i(q^A, p^i, \dot q^\alpha), ~ 
\Phi_\alpha\equiv p_{\alpha}-f_{\alpha}(q^A, p_j)=0$.  
By definition, Hamiltonian formulation of the theory (\ref{201}) is the following system  
on extended phase space with the coordinates $(q^A, p_A, v_\alpha)$:  
$\dot q^A=\{q^A, H\}, ~  \dot p_A=\{p_A, H\}, ~ 
\Phi_\alpha(q^A, p_B)=0$, 
where $\{ ~ , ~ \}$ is the Poisson bracket, and Hamiltonian has the structure 
\begin{eqnarray}\label{206}
H(q^A, p_A, v^\alpha)=H_0(q^A, p_j)+v^\alpha\Phi_\alpha(q^A, p_B).
\end{eqnarray}
%\begin{eqnarray}\label{207}
%H_0=\left(p_i\dot q^i-L+ \dot q^\alpha\frac{\partial\bar 
%L}{\partial \dot q^\alpha}\right) 
%\Biggr|_{\dot q^i\rightarrow v^i(q^A, p_j, \dot q^\alpha)}.
%\end{eqnarray}
The variables $v^{\alpha}$ are called Lagrangian multipliers to the primary constraints. 
It is known [11] that the formulations (\ref{201}) and (\ref{206}) are equivalent.  
Second stage of the Dirac procedure consist in analysis of 2-stage equations 
$\{\Phi_{\alpha}, H\}=0$, the latter are algebraic consequences of the Hamiltonian equations.
Some of the equations can be used for determining of a subgroup of Lagrangian multipliers 
in an algebraic way. Among the remaining equations one takes functionally independent subsystem, the latter 
represent secondary Dirac constraints $\Phi_{\alpha_2}(q^A, p_j)=0$. They imply third-stage equations, 
an so on. We suppose that the theory has 
constraints up to at most $N$ stage: $\Phi_{\alpha}, \Phi_{\alpha_2}, \ldots , \Phi_{\alpha_N}$. 
  
Using these notations, main result of our work can be described schematically as follows. 
Let $\Phi_{\alpha_{s-1}}$ 
be constraints of $(s-1)$-stage of the Dirac procedure. On the next stage one studies the equations
$\{\Phi_{\alpha_{s-1}}, H\}=0$  
for revealing of $s$-stage constraints. Under the above mentioned restriction, {\sf $s$-stage Dirac functions} 
$\{\Phi_{\alpha_{s-1}}, H\}$ can be rewritten in the form  
\begin{eqnarray}\label{101}
\{\Phi_{\alpha_{s-1}}, H\}=A 
\left(
\begin{array}{c}
\pi^{(s)}{}_i(v)\\
\Phi_{\alpha_s}+B_{\alpha_s}(\pi^{(s-1)}, \ldots , \pi^{(2)})\\
C_a(\Phi_{\alpha_{s-1}}, \ldots ,\Phi_{\alpha_2})+
D_a(\pi^{(s-1)}, \ldots ,\pi^2).
\end{array}
\right)
\end{eqnarray}
In particular, the representation is true for general quadratic theory. Here 
$\pi^{(s)}{}_i(v)=0$ represent equations for determining of the Lagrangian multipliers of these stage,  
$\Phi_{\alpha_s}$ are $s$-stage constraints, and $B, C, D$ are linear homogeneous functions of 
indicated variables, with coefficients dependent on $q^A, p_j$. The matrix $A$ and matrices which form 
$B, C, D$ will be called {\sf $s$-stage structure matrices}. 
It may happens that some components of the column 
(\ref{101}), namely "$a$-components", do not represent independent restrictions on the variables $(q, p, v)$. 
The number $[a]$ of these components will be called {\sf defect of $s$-stage system}. Then  
$[a]$ independent local symmetries of the Lagrangian action can be constructed   
$\delta q^A=
\sum^{s-2}_{p=0}{\stackrel{(p)}{\epsilon}{}^a}
R^{(p)}_a{}^{A}(q, \dot q)$,
where generators $R$ are specified in terms of the structure matrices in an algebraic way. 
As it will be shown in Sect.5, the head of the chain $R^{(s-2)}$ has simple algebraic interpretation 
as a projector on "$a$-subspace" of the column (\ref{101}). We present 
manifest form of the symmetries in terms of the structure matrices (see Eq.(\ref{909})  
for s-stage symmetries and Eqs.(\ref{906})-(\ref{908}) for lower-stage symmetries).
 
Thus, knowledge of a structure of $s$-stage Dirac functions (\ref{101}) is equivalent to knowledge 
of $s$-stage local symmetries.

The work is organised as follows. In Sect. 2 we present the so called generating equations in terms of 
Hamiltonian quantities. The equations turn out to be sufficient conditions for existence of local symmetry 
of Lagrangian action. The statement presented in this section is true for general Lagrangian theory. 
Search for solutions 
of the equations implies detailed analysis of $s$-stage Dirac functions. So, in Sect. 3 we demonstrate 
that the Dirac functions can be identically rewritten in the {\sf normal form} (\ref{101}). Using the 
normal form, we analyse the generating equations in Sect. 4. In Sect. 5 we obtain algebraic solution of the 
generating equations for the case of general quadratic theory.  
Proof of some statements is omitted and can be find in [13].

\section{Generating equations for gauge symmetry generators} 

Let us consider infinitesimal transformation 
\begin{eqnarray}\label{301}
q^A\longrightarrow q^{'A}=q^A+\delta q^A, \qquad
\delta q^A=
\sum^{s-2}_{p=0}{\stackrel{(p)}{\epsilon}}
R^{(p)}{}^A(q, \dot q, \ddot q, ...),
\end{eqnarray}
where {\sf parameter} $\epsilon(\tau)$ is arbitrary function of time $\tau$, 
and it was denoted
$\stackrel{(p)}{\epsilon}\equiv\frac{d^p}{d\tau^p}\epsilon$. The transformation is 
{\sf local (or gauge) symmetry}  
of an action $S$, if it leaves $S$ invariant up to surface term 
$\delta L=\frac{d}{d\tau}\omega$,
with some functions $\omega(q, \epsilon)$. 
Local symmetry implies appearance of 
identities among equations of motion of the theory. For a theory without higher derivatives and 
{\sf generators} of the form 
%\footnote{Analysis of this work shows that gauge generators can be find in this form.} 
$R^{(p){}A}(q, \dot q)$, the identities were analyzed in some 
details in our work [10]. First order form (that is the identities on configuration-velocity space), 
and then 
Hamiltonian form of the identities have been obtained. Necessary and sufficient 
conditions for existence of local symmetry of the Lagrangian action can be formulated on this ground. 
Namely, the Hamiltonian identities can be considered as a system of partial differential equations  
for the Hamiltonian counterparts of the functions $R^{(p)}{}^A(q, \dot q)$. 
The equations has been obtained in [10] starting from 
hypothesis that the action is invariant, and by substitution of the velocities $v^i(q^A, p_j, v^\alpha)$  
into the 
first order identities, that is as necessary conditions for existence of gauge symmetry. 
As it was explained in [12], this substitution is change of variables 
on configuration-velocity space, which implies that the resulting Hamiltonian equations 
represent sufficient conditions also. In the present work 
we propose pure algebraic procedure to solve these equations. So, let us present 
the sufficient conditions in a form convenient for subsequent analysis. 

For given integer number $s$, let us construct {\sf generating functions} $T^{(p)}, ~ p=2,3, \ldots ,s$ 
according to the recurrence relations ($T^{(1)}=0$)
\begin{eqnarray}\label{303}
T^{(p)}=Q^{(p) \alpha}\{\Phi_\alpha, H\}+\{H, T^{(p-1)}\},
\end{eqnarray}
where $Q^{(p) \alpha}(q^A, p_j, v^\alpha)$ are some functions. Then one can prove  
the following 

\noindent
{\bf{Statement 1.}} Let the coefficients $Q^{(p) \alpha}, p=2,3, \ldots ,s$ have been chosen in such a way that 
the following {\sf generating equations}:
\begin{eqnarray}\label{304}
\frac{\partial}{\partial v^\alpha}T^{(p)}=0, \quad
p=2,3 \ldots ,s-1; \qquad \qquad 
T^{(s)}=0, 
\end{eqnarray}
hold. Using these $Q$, let us construct the Hamiltonian functions $R^{(p) A}(q^A, p_j, v^\alpha)$, 
$p=0,1,2, \ldots ,s-2$
\begin{eqnarray}\label{306}
R^{(p) \alpha}=Q^{(s-p) \alpha}, \qquad
R^{(p) i}=\{q^i, \Phi_\alpha\} R^{(p)\alpha}-\{q^i, T^{(s-1-p)}\}, 
\end{eqnarray}
and then the Lagrangian functions
\begin{eqnarray}\label{307}
R^{(p) A}(q, \dot q)\equiv
R^{(p) A}(q^A, p_j, v^\alpha)\Biggr|_{p_j\rightarrow\frac{\partial\bar L}{\partial v^j}}
\Biggr|_{v^A\rightarrow\dot q^A}.
\end{eqnarray}
Then the transformation (\ref{301}) is local symmetry of the Lagrangian action. 

Some relevant comments are in order.

\noindent
1) From Eq.(\ref{306}) it follows that only $R^{(p) \alpha}$-block of the Hamiltonian generators is 
essential quantity. On this reason, only this block will be discussed below.

\noindent
2) Hamiltonian generators (\ref{306}) can be used also to construct a local symmetry of Hamiltonian 
action. Expressions for the transformations $\delta q^A, \delta p_A, \delta v^\alpha$ can be 
find in [10]. Let us point that construction of Lagrangian and Hamiltonian symmetries are not exactly 
equivalent tasks, see [10]. Our statement gives sufficient conditions for both symmetries.  

\noindent
3) Search for the symmetry (\ref{301}) (the latter involves derivatives of the parameter 
$\epsilon$ up to order $s-2$ )  
is directly related with $s$-stage of the Dirac procedure. Actually,  
in Sect. 4 we demonstrate that the coefficients $Q^{(p)},~ p=2, 3, \ldots ,s$ 
can be chosen in such a way that each generating function $T^{(p)}$ is linear combination of constraints 
$\Phi_{\alpha_k}$ of the stages $k=2, 3, \ldots ,p$. In particular, 
$T^{(s-1)}=\sum^{s-1}_{p=2}{c^{\alpha_p}\Phi_{\alpha_p}}$, then 
$T^{(s)}$ in Eq.(\ref{304})
involves the Dirac functions up to $s$-stage: $T^{(s)}\sim\{T^{(s-1)}, H\}\sim$
$\{\Phi_{\alpha_p}, H\},~ p=2, 3, \ldots ,s-1$.  
Then the symmetry (\ref{301}) can be called {\sf $s$-stage symmetry}, while   
$T^{(2)}, T^{(3)}, \ldots ,T^{(s)}$ 
will be called {\sf $s$-stage generating functions}. 

\noindent
4) According to the statement, symmetries of different stages $s$ can be looked for separately. 
To find 2-stage 
symmetries $\delta_{a_2} q^A=\epsilon^{a_2} R_{a_2}^{(0) A}$, one look for solutions $Q_{a_2}^{(2) \alpha}$ 
of the 
equation 
\begin{eqnarray}\label{309}
T^{(2)}\equiv Q^{(2) \alpha}\{\Phi_\alpha, H\}=0.
\end{eqnarray}
Note that it implies analysis of second-stage Dirac functions $\{\Phi_\alpha, H\}$. 
3-stage symmetries 
$\delta_{a_3} q^A=\epsilon^{a_3} R_{a_3}^{(0) A}+\dot\epsilon^{a_3} R_{a_3}^{(1) A}$
are constructed from solutions $Q_{a_3}^{(2) \alpha}, Q_{a_3}^{(3) \alpha}$ 
of the equations (coefficients $Q$ of different stages are independent) 
\begin{eqnarray}\label{310}
\frac{\partial}{\partial v^\beta}
T^{(2)}\equiv \frac{\partial}{\partial v^\beta}(Q^{(2) \alpha}\{\Phi_\alpha, H\})=0, \quad 
T^{(3)}\equiv Q^{(3) \alpha}\{\Phi_\alpha, H\}+
\left\{Q^{(2)\alpha}\{\Phi_\alpha, H\}, H\right\}=0,
\end{eqnarray}
and so on. In a theory with at most $N$-stage Dirac constraints presented, the procedure stops for 
$s=N+1$, see Sect. 5 below. 

\noindent
5) Since the generating equations (\ref{304}) do not involve the momenta $p_\alpha$, one can 
search for solutions in the form $Q^{(p) \alpha}(q^A, p_j, v^\alpha)$. As a result, Hamiltonian generators 
do not depend on $p_\alpha$. In this case, passage to the Lagrangian first order formulation 
is change of variables [12]: $(q^A, p_i, v^\alpha)\leftrightarrow(q^A, v^i, v^\alpha)$. This change 
has been performed in Eq. (\ref{307}). 

\section{Normal form of $p$-stage Dirac functions}

As it was discussed in the previous section, expression for generating function $T^{(k)}$ involve the  
Dirac functions $\{\Phi_{\alpha_p}, H\}$, $p=1, 2, \ldots ,k-1$. One needs to know detailed 
structure of them to solve the generating equations. Let us point that 
this part of analysis is, in fact, part of the Dirac procedure for revealing of higher-stage 
constraints. The only difference is that in the Dirac procedure one studies the equations   
$\{\Phi_{\alpha_p}, H\}=0$, where constraints and equations for Lagrangian multipliers 
of previous stages can be used. Since our generating equations must be satisfied by $Q$ for 
any $q, p, v$, 
one needs now to study the Dirac functions outside of extremal surface. Below we suppose that 
matrices of the type $\{\Phi_{\alpha_p}, \Phi_{\alpha}\}$ have constant rank in vicinity of 
phase space point under consideration. In particular, it is true for quadratic theory.
In this section we describe an induction 
procedure to represent $p$-stage Dirac functions in the normal form convenient for 
subsequent analysis, see Eq.(\ref{601}) below. On each stage, it will be necessary to 
divide some groups of functions on subgroups. Let us start with detailed analysis of second stage, with 
the aim to clarify notations which will be necessary to work out $p$-stage Dirac functions and the 
corresponding generating equations.

\noindent
{\bf Second-stage Dirac functions.} 
With a group of quantities appeared on first stage of the Dirac procedure we assign number 
of the stage, the latter replace corresponding index (the number will be called 
{\sf index of the group} below).  
Then the primary constraints are $\Phi_\alpha\equiv\Phi_1$, and the Lagrangian 
multipliers are denoted as $v^{\alpha}\equiv v^1$. Number of functions in a group is denoted as 
$[1]\equiv [\alpha]$.  
For the second stage Dirac functions one writes
\begin{eqnarray}\label{401}
\{\Phi_\alpha, \Phi_\beta v^\beta +H_0\} \rightarrow
\{\Phi_1, \Phi_{1'} v^{1'} +H_0\}= 
\{\Phi_1, \Phi_{1'}\} v^{1'} +\{\Phi_1, H_0\}\equiv
\triangle_{(2)1 1'}v^{1'}+H_{(2)1}.
\end{eqnarray}
So, repeated up and down number of stage imply summation over the corresponding indices.
With quantities first appeared on second stage has been assigned number of the stage: 
$\triangle_{(2)}, ~ H_{(2)}$ 
(where confusion is not possible, it can be omitted).  
Suppose that $rank\triangle_{(2)1 1'}=\left[\overline{2}\right]$, then one finds 
$\left[\widetilde{2}\right]=[1]-\left[\overline{2}\right]$ 
independent null-vectors $\vec K_{(2) \widetilde{2}}$ of the matrix $\triangle_{(2)}$ with components 
$K_{(2) \widetilde{2}}{}^1$. Let $K_{(2) \overline{2}}{}^1$ be any completion of the set 
$\vec K_{(2) \widetilde{2}}$ up to a basis of $[1]$-dimensional space. By construction, the matrix    
\begin{eqnarray}\label{402}
K_{(2) \widehat 1}{}^1\equiv
\left(K_{(2) \overline{2}}{}^1\atop  K_{(2) \widetilde{2}}{}^1\right),
\end{eqnarray}
is invertible. For any matrix $K$, the inverse matrix is denoted as $\widetilde K$: 
$\widetilde K_{(2)1}{}^{\widehat 1} K_{(2) \widehat 1}{}^{1'}=\delta_1{}^{1'}$. 
The matrix $K$ is a kind of {\sf conversion matrix} which transforms the index $1$ into  
$\widehat 1$, the latter is naturally  divided on two groups   
$1\rightarrow \widehat 1=(\bar 2, \widetilde{2})$. 
Since $K_{(2) \widetilde{2}}{}^1\triangle_{(2)1 1'}=0$, 
the conversion matrix can be used to separate the Dirac functions on $v$-dependent and $v$-independent 
parts
\begin{eqnarray}\label{404}
\begin{array}{c}
\{\Phi_1, H\}=\widetilde K K \{\Phi_1, H\}=
\widetilde K_{(2) 1}{}^{\widehat 1}\left(\pi_{\overline{2}}(v^1)\atop 
\Phi_{\widetilde{2}}(q^A, p_j)\right), \cr
\pi_{\overline{2}}(v^1)\equiv X_{(2)\overline 2 1}v^1+Y_{(2) \overline 2}, \quad  
\Phi_{\widetilde{2}}\equiv K_{(2) \widetilde{2}}{}^1 H_{(2)1}.
\end{array}
\end{eqnarray}
Here it was denoted 
%\begin{eqnarray}\label{405}
$X_{(2)\overline 2 1}=K_{(2) \overline 2}{}^{1'}\triangle_{(2)1' 1}, ~   
Y_{(2) \overline 2}=K_{(2) \overline 2}{}^1 H_{(2)1}$. 
%\end{eqnarray}

Let us analyze the functions $\pi_{\overline{2}}(v^1)$. By construction, the matrix $X$ has maximum rank 
equal $\left[\overline{2}\right]$. Without loss of generality, we suppose that from the beginning $v^1$ 
has been chosen 
such that the rank columns appear on the left: 
$X_{(2)\overline{2} 1}=(X_{(2)\overline{2} \overline{2}}, X_{(2)\overline{2} \underline{2}})$. 
So, the Lagrangian multipliers are divided on two groups 
$v^1=(v^{\overline{2}}, v^{\underline{2}})$, one 
writes\footnote{On this stage one has $[\widetilde{2}]=[\underline{2}]$, but it will not 
be true for higher stages. On this reason we adopt different notations for these groups.}
\begin{eqnarray}\label{406}
\begin{array}{c}
\pi_{\overline{2}}=X_{(2) \overline{2} \overline{2}}v^{\overline{2}}+
X_{(2) \overline{2} \underline{2}}v^{\underline{2}}
+Y_{(2) \overline{2}}, \Longrightarrow 
v^{\overline{2}}\equiv\widetilde X_{(2)}{}^{\overline{2} \overline{2'}}\pi_{\overline{2'}}(v^1)+
\Lambda_{(2)}{}^{\overline{2}}{}_{\underline{2}}v^{\underline{2}}+W_{(2)}{}^{\overline{2}},
\end{array}
\end{eqnarray}
where
%\begin{eqnarray}\label{4077}
$\Lambda_{(2)}{}^{\overline{2}}{}_{\underline{2}}=-\widetilde X_{(2)}{}^{\overline{2} \overline{2'}}
X_{(2) \overline{2'} \underline{2}}, ~ 
W_{(2)}{}^{\overline{2}}=-\widetilde X_{(2)}{}^{\overline{2} \overline{2'}}Y_{(2) \overline{2'}}$.  
%\end{eqnarray}
We stress that the second equation in (\ref{406}) is an identity. It will be necessary to analyze 
higher-stage Dirac functions below. 

Let us analyze the functions $\Phi_{\widetilde{2}}$ in Eq.(\ref{404}). By construction, they depend 
on the phase space variables $z_1\equiv(q^A, p_j)$. According to Dirac, functionally independent functions 
among $\Phi_{\widetilde{2}}$ are called secondary constraints, and the equations $\Phi_{\widetilde{2}}=0$ 
can be used to express a part $\bar z_2$ of the phase space variables $z_1=(\bar z_2, z_2)$  
in terms of $z_2$. Let us suppose 
\begin{eqnarray}\label{408}
rank\frac{\partial\Phi_{\widetilde{2}}}{\partial z_1}\Biggr|_{\Phi_{\widetilde{2}}}=
rank\frac{\partial\Phi_{\widetilde{2}}}{\partial \bar z_2}\Biggr|_{\Phi_{\widetilde{2}}}=[\bar z_2].
\end{eqnarray}
Then $\Phi_{\widetilde{2}}$ can be identically rewritten in the form [13]  
\begin{eqnarray}\label{409}
\Phi_{\widetilde{2}}(z_1)=U_{(2) \widetilde{2}}{}^{\widetilde{2'}}
\left(\Phi_2(z_1)\atop 0_{\breve 2}\right), \qquad 
\left[\Phi_2\right]=[\bar z_2], \qquad \det U\ne 0,
\end{eqnarray}
where index $\widetilde 2$ is divided on two groups $\widetilde 2=(2, \breve 2)$, and $\Phi_2$ 
are functionally independent.

Substitution of this result into Eq.(\ref{404}) gives the normal form of second-stage Dirac functions
\begin{eqnarray}\label{413}
\{\Phi_1, H\}=
A_{(2) 1}{}^{\widehat 1}(q^A, p_j)
\left(
\begin{array}{c}
\pi_{\overline{2}}(v^1)\\
\Phi_2(q^A, p_j)\\
0_{\breve 2}
\end{array}
\right), \qquad 
A_{(2) 1}{}^{\widehat 1}\equiv\widetilde K_{(2)}
\left(
\begin{array}{cc}
1_{(\overline{2})\times(\overline{2})}&0\\
0&U_{(2) \widetilde{2}}{}^{\widetilde{2}}
\end{array}
\right).
\end{eqnarray}
where $A$ is invertible matrix ,  
and functions $\pi_{\overline{2}}(v^1)$ are given by Eq.(\ref{406}). In the process, the Lagrangian 
multipliers $v^1$ have been divided on subgroups $(v^{\overline 2}, v^{\underline 2})$, where 
$v^{\overline 2}$ can be identically rewritten in terms of $v^{\underline 2}, \pi_{\underline 2}$ 
according to Eq.(\ref{406}). The functions 
$\Phi_\alpha=p_\alpha-f_\alpha(q^A, p_j), ~ \Phi_2(q^A, p_j)$ are functionally independent, 
$\Phi_2$ represent all secondary constraints of the theory. By construction, 
$\pi_{\overline 2}(v^1)=0$ are
equations for determining of the multipliers $v^{\overline 2}$.

"Evolution" of the index $1$ of previous stage during the second stage can be 
resumed as follow: 
it can either be divided on two subgroups: $1=(\overline{2}, \underline{2})$, 
or can be converted into $\widehat 1$ and then divided on three subgroups:
$1\rightarrow\widehat 1=(\overline{2}, 2, \breve 2)$. Similar notations are used for higher stages: 
on some stage $p$, index $\underline{p-1}$ of previous stage can be divided on two subgroups 
$\underline{p-1}=(\overline{p}, \underline{p})$. 
Index ${p-1}$ of previous stage can be converted into $\widehat{p-1}$ and then divided on three 
subgroups $p-1\rightarrow\widehat{p-1}=(\overline{p}, p, \breve p)$.
If $k< p-1$, the notations $\psi_{(p) 1}, \psi_{(p) \underline{k}}$ mean 
$\psi_{(p) \underline{k}}$$=(\psi_{(p) \overline{k+1}}, \psi_{(p) \overline{k+2}}, $$\ldots ,$$
\psi_{(p) \overline{p-1}}, \psi_{(p) \underline{p-1}})$.

{\bf $p$-stage Dirac functions.} Similarly to $p=2$ case discussed above, higher stage Dirac functions 
can be identically rewritten in the normal form. Let $\Phi_1, \Phi_2, \ldots , \Phi_{p-1}$ is set of constraints 
and $\pi_{\overline 2}, \pi_{\overline 3}, \ldots , \pi_{\overline{p-1}}$ is set of equations for Lagrangian  
multipliers of previous stages. Then  
\begin{eqnarray}\label{601}
\{\Phi_{p-1}, H\}=
A_{(p) p-1}{}^{\widehat{p-1}} 
\left(
\begin{array}{c}
\pi_{\overline{p}}(v)\\
\Phi_p(q^A, p_j)+B_{(p) p}(\pi_{\overline{p-1}}, \ldots ,\pi_{\overline{2}})\\
C_{(p)\breve p}(\Phi_{p-1}, \ldots ,\Phi_2)+
D_{(p)\breve p}(\pi_{\overline{p-1}}, \ldots ,\pi_{\overline 2})
\end{array}
\right), 
\end{eqnarray}
where the matrix $A_{(p) p-1}{}^{\widehat{p-1}}(z_1)$ is invertible, $\Phi_p$ are functionally independent and can be identified with $p$-stage constraints, $\Phi_k$, $k=1, 2, \ldots , p$ are 
functionally independent functions also (note that the primary constraints are included). 
$B, C, D$ are linear homogeneous functions of indicated variables, for example 
$C_{(p)\breve p}=\sum_{k=2}^{p-1} C_{(p)\breve p}{}^k(z_1)\Phi_k$.
Proof can be done by induction over number of stage $p$, and is presented in [13].  
It is easy to see that our definition of $p$-stage constraints is equivalent to the 
standard one. Division 
on subgroups has been made in accordance with properties of the system (\ref{601}),   
which determines dimensions of the subgroups. In particular, 
$\left[\breve p\right]=\left[p-1\right]-\left[\overline{p}\right]-[p]$ is called {\sf{defect}} of 
$p$-stage Dirac system ({\ref{601}). Number of independent (but possibly reducible) $p$-stage 
symmetries, which can be find by our procedure, coincides with the defect $\left[\breve p\right]$, 
see below.

\section{Normal form of generating functions $T^{(p)}$}

Here we describe procedure to rewrite the set of $s$-stage generating functions 
(\ref{303}) in the {\sf normal form} (\ref{802}), that is as combination of constraints. Let us start from 
analysis of lower stages. 

\noindent
{\bf Normal form of second-stage generating function.} Using Eq.(\ref{413}) one immediately obtains the desired 
result
\begin{eqnarray}\label{701}
T^{(2)}=Q^{(2)1}\{\Phi_1, H\}=-Q^{(2) 2}\Phi_2,
\end{eqnarray}
where the coefficients $Q^{(2) 1}(q^A, p_j)$ have been chosen as follow 
\begin{eqnarray}\label{703}
\begin{array}{c}
Q^{(2)1}\equiv\widehat Q^{(2) \widehat 1}\widetilde A_{(2) \widehat 1}{}^{1}=(\widehat Q^{(2) \overline 2},   
\widehat Q^{(2) 2}, \widehat Q^{(2) \breve 2})\widetilde A_{(2) \widehat 1}{}^{1}, \qquad  
\widehat Q^{(2) \overline 2}=0, \qquad  
\widehat Q^{(2) 2}=-Q^{(2) 2}, 
\end{array} 
\end{eqnarray}
and $Q^{(2) 2}, ~ \widehat Q^{(2) \breve 2}$ remain arbitrary. 
Taking further 
$Q^{(2) 2}=0$, one obtains solution (\ref{701}), (\ref{703}) of second-stage generating equation (\ref{309}). 
According to Statement 1, one writes out immediately $\left[\breve 2\right]$ independent local 
symmetries of the Lagrangian action, see Sect. 5. Number of second-stage symmetries coincides with 
the defect of second-stage system (\ref{413}). 
The symmetries are specified by second-stage structure matrix $A_{(2)}$.

\noindent
{\bf{Normal form of third-stage generating functions.}} One has the set
\begin{eqnarray}\label{706}
T^{(2)}=Q^{(2)1}\{\Phi_1, H\}, \qquad 
T^{(3)}=Q^{(3) 1}\{\Phi_1, H\}+\{H, T^{(2)}\},
\end{eqnarray}
where the coefficients $Q^{(2) 1}(q^A, p_j, v^{\alpha}), ~ Q^{(3) 1}(q^A, p_j, v^{\alpha})$ are 
some functions.   
As in the previous case, making the choice (\ref{703}), one writes $T^{(2)}$ in the normal 
form (\ref{701}). Using this expression for $T^{(2)}$, as well as Eq.(\ref{601}), one obtains the following 
expression for $T^{(3)}$ 
\begin{eqnarray}\label{708}
\begin{array}{c}
T^{(3)}=Q^{(3) 1}\{\Phi_1, H\}-\{H, Q^{(2) 2}\Phi_2\}= \cr 
\left(\widehat Q^{(3) \overline 2}+\widehat Q^{(2) 3}B_{(3) 3}{}^{\overline 2}+
\widehat Q^{(2) \breve 3}D_{(3)\breve 3}{}^{\overline 2}\right)\pi_{\overline{2}}+
\widehat Q^{(2) \overline 3}\pi_{\overline{3}}+ \cr 
\left(\widehat Q^{(3) 2}-\{H, Q^{(2)2}\}+\widehat Q^{(2) \breve 3}C_{(3)\breve 3}{}^2\right)\Phi_2+
\widehat Q^{(2) 3}\Phi_3+
\widehat Q^{(3) \breve 2}0_{\breve 2}.
\end{array}
\end{eqnarray}
where 
\begin{eqnarray}\label{709}
Q^{(2) 2}A_{(3) 2}{}^{\widehat 2}\equiv\widehat Q^{(2) \widehat 2}=(\widehat Q^{(2) \overline 3}, ~  
\widehat Q^{(2) 3}, ~ \widehat Q^{(2) \breve 3}), \quad  
Q^{(3)1}A_{(2) 1}{}^{\widehat 1}\equiv\widehat Q^{(3) \widehat 1}=(\widehat Q^{(3) \overline 2}, ~  
\widehat Q^{(3) 2}, ~ \widehat Q^{(3) \breve 2}).  
\end{eqnarray}
Then the following choice 
\begin{eqnarray}\label{711}
\widehat Q^{(2) \overline 3}=0, ~  
\widehat Q^{(3) \overline 2}=-\widehat Q^{(2) 3}B_{(3) 3}{}^{\overline 2}-
\widehat Q^{(2) \breve 3}D_{(3)\breve 3}{}^{\overline 2}, \qquad \quad \\[1pt] \nonumber
\widehat Q^{(2) 3}\equiv -Q^{(2) 3}, ~  
\widehat Q^{(3) 2}=-Q^{(3) 2}+\{H, \widehat Q^{(2) \widehat 2}\widetilde A_{(3) \widehat 2}{}^2\}-
\widehat Q^{(2) \breve 3}C_{(3)\breve 3}{}^2,
\end{eqnarray}
with arbitrary functions $Q^{(2) 3}, ~ Q^{(3) 2}$, gives $T^{(3)}$ in the normal form:   
$T^{(3)}=-Q^{(3) 2}\Phi_2-Q^{(2) 3}\Phi_3$. 
Thus the normal form for third-stage generating functions is supplied by  
special choice of $Q^{(2) 1}, ~ Q^{(3) 1}$. The coefficients have been divided on the following groups: 
\begin{eqnarray}\label{7133}
Q^{(2) 1}=\left(0^{\overline 2}, \left(0^{\overline 3}, \widehat Q^{(2) 3}, 
\widehat Q^{(2) \breve 3}\right)\widetilde A_{(3) \widehat 2}{}^2, \widehat Q^{(2) \breve 2}\right)
\widetilde A_{(2) \widehat 1}{}^1, \cr
Q^{(3) 1}=\left(\widehat Q^{(3) \overline 2}, \widehat Q^{(3) 2}, \widehat Q^{(3) \breve 2}\right)
\widetilde A_{(2) \widehat 1}{}^1. \qquad \qquad
\end{eqnarray}
To describe structure of the groups, it is convenient to use the following triangle table 
\begin{eqnarray}\label{713}
\begin{array}{ccccccccccc}
Q^{(2) 1}&\sim&&&&0^{\overline 2}&0^{\overline 3}&Q^{(2) 3}&\widehat Q^{(2) \breve 3}&
\widehat Q^{(2) \breve 2}\\
Q^{(3) 1}&\sim&&&&&\widehat Q^{(3) \overline 2}&Q^{(3) 2}&\widehat Q^{(3) \breve 2}&\\
\end{array}
\end{eqnarray}
Writting out such a kind  
tables below, we omite the conversion matrices and write arbitrary coefficients $Q^k$ instead of 
$\widehat Q^k$ in central column of the table.
Then the central column and columns on r.h.s. of it represent coefficients which remains arbitrary on 
this stage.   
The coefficients $Q^{(3) 2}=Q^{(2) 3}=0$ appeared in $T^{(3)}$ remains arbitrary functions, 
while $Q^{(2) 2}$ in $T^{(2)}$ is 
$Q^{(2) 2}=$$\left(0^{\overline 3}, -Q^{(2) 3}, 
\widehat Q^{(2) \breve 3}\right)\widetilde A_{(3) \widehat 2}{}^2$.  
Taking further  
$Q^{(3) 2}=Q^{(2) 3}=0$, one obtains solution of third-stage generating 
equations (\ref{310}). 
According to Statement 1, it implies $\left[\breve 3\right]$ independent third-stage local 
symmetries of the Lagrangian action, see Sect. 5. Number of the symmetries coincides with the defect 
of third-stage Dirac system (\ref{601}). 
The symmetries are specified, in fact, by the structure matrices 
$A_{(2)}, A_{(3)}, B_{(3)}, C_{(3)}, D_{(3)}$. 

{\bf Normal form of $s$-stage generating functions.} Now it is clear that induction over number of stage 
$s$, $2\le s\le N+1$ can be used to rewrite the set: $T^{(2)}, T^{(3)}, \ldots ,T^{(s-1)},T^{(s)}$ 
in the normal form. Supposing that $T^{(2)}, T^{(3)}, \ldots ,T^{(s-1)}$ have been presented already 
in the normal form, and following the same line as before, one proves [13] the following    

{\bf Statement 2.} Consider the Lagrangian theory with the Hamiltonian $H$, and with constraints 
at most $N$ stage appeared in the Hamiltonian 
formulation: $\Phi_1, \Phi_2, \ldots ,\Phi_N$. For some fixed integer number $s$, $2\le s\le N+1$, let 
us construct set of generating 
functions according to recurrence relations ($T^{(1)}=0$):
$T^{(p)}=Q^{(p) 1}\{\Phi_1, H\}+\{H, T^{(p-1)}\}, \qquad p=2,3, \ldots ,s$. 
Then the coefficients $Q^{(p)1}(q^A, p_j, v^{\alpha})$ can be chosen in such a way, that 
all $T^{(p)}$ turn out to be linear combinations of the constraints  
\begin{eqnarray}\label{802}
T^{(p)}=-\sum_{n=2}^p Q^{(p+2-n) n}\Phi_n, \qquad p=2,3, \ldots ,s. 
\end{eqnarray}
Choice of $Q^{(p)1}$, which supplies the normal form, can be described as follows: \par
\noindent
a) For any $k=2,3, \ldots ,s, ~ n=1,2, \ldots ,(s+1-k)$, $Q^{(k) n}$ is divided on three subgroups  
with help of the structure matrix of $(n+1)$-stage $A_{(n+1)}$ 
\begin{eqnarray}\label{8033}
Q^{(k) n}A_{(n+1) n}{}^{\widehat n}\equiv\widehat Q^{(k) \widehat n}=
(\widehat Q^{(k) \overline{n+1}}, \widehat Q^{(k) n+1}, \widehat Q^{(k) \breve{n+1}}), 
\end{eqnarray}
where for any $k=2,3, \ldots ,s, ~ n=2,3, \ldots ,(s+2-k)$
\begin{eqnarray}\label{803}
\begin{array}{c}
\widehat Q^{(k) n}=-Q^{(k) n}+\{H, \widehat Q^{(k-1) \widehat n}
\widetilde A_{(n+1) \widehat n}{}^n\}-
\sum_{m=2}^{k-1} \widehat Q^{(m) \breve{k+n-m}}C_{(k+n-m) \breve{k+n-m}}{}^n, 
\end{array} 
\end{eqnarray}
\begin{eqnarray}\label{804}
\begin{array}{c}
\widehat Q^{(k) \overline n}=-
\sum_{m=2}^{k-1}\left(\widehat Q^{(m) k+n-m}B_{(k+n-m) k+n-m}{}^{\overline n}+\right.
\left. \widehat Q^{(m) \breve{k+n-m}}D_{(k+n-m) \breve{k+n-m}}{}^{\overline n}\right), 
\end{array}  
\end{eqnarray}
and $B, ~ C, ~ D$ are structure matrix of the Dirac procedure, see Eq.(\ref{601}). Eqs.(\ref{803}), (\ref{804}) 
imply $\widehat Q^{(2) p}=-Q^{(2) p}, ~ \widehat Q^{(2) \overline p}=0$.\par
\noindent
b) The coefficients $\widehat Q^{(k) \breve{n+1}}, ~ k=2,3, \ldots ,s, ~ n=1,2, \ldots ,(s+1-k)$ remain 
arbitrary. \par
\noindent
c) The coefficients $Q^{(s+2-n) n}, ~ n=2,3, \ldots ,s$ remain arbitrary.
  
Let us confirm that the recurrence relations (\ref{803}), (\ref{804}) 
actually determines the coefficients.
According to the statement, $Q^{(p)1}$ is converted into $\widehat Q^{(p) \widehat 1}$, and then is 
divided on subgroups 
$(\widehat Q^{(p) \overline 2}, \widehat Q^{(p) 2}, \widehat Q^{(p) \breve 2})$. Then  
$Q^{(p) 2}$ is picked out from $\widehat Q^{(p) 2}$ according to Eq.(\ref{803}). The coefficient 
$Q^{(p) 2}$ will be further converted and divided, creating $Q^{(p) 3}$, and so on. Resulting expression 
for $Q^{(p)1}, p=2, 3, \ldots , s$ on the stage $s$ is 
\begin{eqnarray}\label{3.10}
\begin{array}{l}
Q^{(p) 1}=\left(\widehat Q^{(p) \overline 2}, \left(\widehat Q^{(p) \overline 3},
\left(\left.\ldots\left(\widehat Q^{(p) \overline{s+2-p}}, 
\widehat Q^{(p) s+2-p} 
\widehat Q^{(p) \breve{s+2-p}}\right)^{\widehat{s+1-p}}\right.\right.\right.\right. \cr
\left.\left.\left.\left.\widetilde A_{(s+2-p) \widehat{s+1-p}}{}^{s+1-p}, 
\widehat Q^{(p) \breve{s+1-p}}\right)^{\widehat{s-p}}\widetilde A_{(s+1-p) \widehat{s-p}}{}^{s-p},
\widehat Q^{(p) \breve{s-p}}\right)^{\widehat{s-p-1}}\ldots , \right.\right. \cr 
\left.\left.\widehat Q^{(p) \breve 3}\right)^{\widehat 2}\widetilde A_{(3) \widehat 2}{}^2, 
\widehat Q^{(p) \breve 2}\right)^{\widehat 1}\widetilde A_{(2) \widehat 1}{}^1. 
\end{array}
\end{eqnarray}
Structure of the coefficients $Q^{(p)1}$ can be described by the following table (first line 
represents $Q^{(2)1}$, second line represents $Q^{(3)1}$, and so on, up to $Q^{(s)1}$): 
\begin{eqnarray}\label{805}
\begin{array}{ccccccccccc}
0^{\overline 2}&0^{\overline 3}&0^{\overline 4}&\ldots&0^{\overline s}&
Q^{(2) s}&\widehat Q^{(2) \breve s}&\ldots&\widehat Q^{(2) \breve 4}&\widehat Q^{(2) \breve 3}&
\widehat Q^{(2) \breve 2}\\
&\widehat Q^{(3)\overline 2}&\widehat Q^{(3)\overline 3}&\ldots&\widehat Q^{(3) \overline{s-1}}&
Q^{(3) s-1}&\widehat Q^{(3) \breve{s-1}}&\ldots&\widehat Q^{(3) \breve 3}&\widehat Q^{(3) \breve 2}&\\
&&\ldots&\ldots&\ldots&\ldots&\ldots&\ldots&\ldots&&\\
&&\widehat Q^{(p)\overline 2}&\ldots&\widehat Q^{(p)\overline{s+2-p}}&Q^{(p) s+2-p}&
\widehat Q^{(p) \breve{s+2-p}}&\ldots&\widehat Q^{(p) \breve 2}&&\\
&&&\ldots&\ldots&\ldots&\ldots&\ldots&&\\
&&&\widehat Q^{(s-1)\overline 2}&\widehat Q^{(s-1)\overline 3}&Q^{(s-1) 3}&
\widehat Q^{(s-1) \breve 3}&\widehat Q^{(s-1) \breve 2}&&&\\
&&&&\widehat Q^{(s)\overline 2}&Q^{(s) 2}&\widehat Q^{(s) \breve 2}&&&&\\
\end{array} 
\end{eqnarray}
Note that any group $Q^{(p)n}$ with $n\ne s+2-p$ is presented on the table by interval of $p$-line between 
$\widehat Q^{(p) \overline{n+1}}$ and $\widehat Q^{(p) \breve{n+1}}$. 

From Eqs.(\ref{803}), (\ref{804}) it follows that any group 
$\widehat Q^{(k) \overline n}$ of the line $k$ of the triangle is expressed through some groups placed in 
previous lines on r.h.s. of $\widehat Q^{(k) \overline n}$.  
Any group $\widehat Q^{(k) n}$ is presented through the interval 
$\left[\widehat Q^{(k) \overline{n+1}}, ~ \widehat Q^{(k) \breve{n+1}}\right]$ of the line $k$ 
as well as through some groups of previous lines placed on r.h.s. of $n$ column. After all, all the 
coefficients are expressed through $Q^{(s+2-n) n}\equiv Q^{(n) s+2-n}$, $\widehat Q^{(k) \breve{n}}$,    
which remain arbitrary function (the latter are placed in the central column and on r.h.s. of it in 
the triangle). 
Note that all arbitrary coefficients $Q^{(s+2-n) n}$ 
appear in the expression for higher generating function $T^{(s)}$. 

Lower-stage generating functions have been described before. Using the statement 2, 4-stage generating 
functions can be described as follow.

\noindent
{\bf{Normal form of 4-stage generating functions.}}
\begin{eqnarray}\label{819}
\begin{array}{l}
T^{(2)}=-Q^{(2) 2}\Phi_2, \quad 
T^{(3)}=-Q^{(3) 2}\Phi_2-Q^{(2) 3}\Phi_3, \quad  
T^{(4)}=-Q^{(4) 2}\Phi_2-Q^{(3) 3}\Phi_3-Q^{(2) 4}\Phi_4.
\end{array}
\end{eqnarray}
The coefficients $Q$ in $T^{(4)}$ remain arbitrary functions, while $Q$ in $T^{(3)}, T^{(4)}$ are
\begin{eqnarray}\label{823}
\begin{array}{c}
Q^{(2) 2}=\left(0^{\overline 3}, \left(0^{\overline 4}, -Q^{(2) 4},
\widehat Q^{(2) \breve 4}\right)\widetilde A_{(4) \widehat 3}{}^3, 
\widehat Q^{(2) \breve 3}\right)\widetilde A_{(3) \widehat 2}{}^2, \cr 
Q^{(2) 3}=\left(0^{\overline 4}, -Q^{(2) 4},
\widehat Q^{(2) \breve 4}\right)\widetilde A_{(4) \widehat 3}{}^3, \cr
Q^{(3) 2}=\left(Q^{(2) 4}B_{(4) 4}{}^{\overline 3}-
\widehat Q^{(2) \breve 4}D_{(4)\breve 4}{}^{\overline 3}, -Q^{(3) 3}+ \right.\cr
\left.\{H, \left(0^{\overline 4}, -Q^{(2) 4}, 
\widehat Q^{(2) \breve 4}\right)\widetilde A_{(4) \widehat 3}{}^3\}-
\widehat Q^{(2) \breve 4}C_{(4) \breve 4}{}^3,
\widehat Q^{(3) \breve 3}\right)\widetilde A_{(3) \widehat 2}{}^2.
\end{array}
\end{eqnarray}
All $Q$ on r.h.s. of these equations are arbitrary functions.

\section{Gauge symmetries of quadratic theory}

Suppose that in the Hamiltonian formulation of our theory there are appear constraints up to at 
most $N$-stage. According to the Statement 1, symmetries of different stages are looked for separately. 
Generators of $s$-stage local symmetries (\ref{306})-(\ref{307}) can be constructed starting from any 
solution of generating equations (\ref{304}). Using normal form (\ref{802})-(\ref{804}) of 
generating functions, 
one concludes that $T^{(s)}=0$ is satisfied by taking $Q^{(p) s+2-p}=0, ~ p=2, 3, \ldots ,s$, i.e. all 
the coefficients in the central column of the triangle (\ref{805}) must be zeros. First equation in 
(\ref{304}) states that generating functions $T^{(p)}$ with $p=2, 3, \ldots ,s-1$ do not depend on 
the Lagrangian multipliers. Dependence on $v^1$ can appear only due to second term in Eq.(\ref{802}).
Thus one needs to kill this term, which can be easily achieved in 
a theory with all the structure matrices $\widetilde A$ (see Eq.(\ref{601})) being numerical matrices.  
It happens, in particular, in any quadratic theory (then all the structure matrices 
$A, B, C, D$ in Eq.(\ref{601}) turn out to be numerical matrices). 
We analyse this case in the present section. For the case, it is consistent to look for solutions with 
$Q=const$, then the second term in Eq.(\ref{803}) disappears, and the generating equations 
(\ref{304}) are trivially satisfied. 

Thus for any quadratic Lagrangian theory it is sufficient to take elements of central column of 
the triangle (\ref{805}) be zeros, and elements on r.h.s. of it be arbitrary numbers, to obtain some 
local symmetry of the Lagrangian action (\ref{301}), (\ref{307}). 

Let us discuss particular set of generators constructed as follows. On stage $s$ of the 
Dirac procedure, one takes $\left[\breve {s'}\right]$ sets of  $\widehat Q^{(2)\breve s}$, namely 
$\widehat Q^{(2)}{}_{\breve {s'}}{}^{\breve s}=\delta_{\breve {s'}}{}^s$, where 
$\left[\breve{s'}\right]$ is defect of 
the system (\ref{601}). Remaining arbitrary coefficients on r.h.s. of the triangle (\ref{805}) 
are taken vanishing. Then these $\left[\breve{s'}\right]$ solutions 
$Q^{(p)}{}_{\breve{s'}}{}^{1}, p=2, 3, \ldots ,s$ of generating equations have the form (\ref{3.10}), 
where one needs to substitute 
$Q^{(p) s+2-p}=0, \qquad \widehat Q^{(2)}{}_{\breve{s'}}{}^{\breve s}=\delta_{\breve{s'}}{}^{\breve s}, ~  
\widehat Q^{(p) \breve k}=0, \qquad p\ne 2, \qquad k\ne s$, 
while others coefficients can be find from Eqs.(\ref{802}), (\ref{803}), the latters acquire the form
$(k=3, 4, \ldots ,s, ~ n=2, 3, \ldots ,s+2-k)$
\begin{eqnarray}\label{902}
\begin{array}{l}
\widehat Q^{(2)}{}_{\breve{s'}}{}^n=-Q^{(2)}{}_{\breve{s'}}{}^n, \qquad
\widehat Q^{(k)}{}_{\breve{s'}}{}^n=-Q^{(k)}{}_{\breve{s'}}{}^n-\delta_{s}{}^{k+n-2}
C_{(s) \breve{s'}}{}^n, \cr
\widehat Q^{(k)}{}_{\breve{s'}}{}^{\overline n}=\sum_{m=2}^{k-1} 
Q^{(m) k+n-m}B_{(k+n-m) k+n-m}{}^{\overline n}- \cr 
\delta_{s}{}^{k+n-2}
\left(D_{(s) \breve{s'}}{}^{\overline n}-\sum_{m=3}^{k-1} C_{(s) \breve{s'}}{}^{k+n-m}
B_{(k+n-m) k+n-m}{}^{\overline n}\right).
\end{array}
\end{eqnarray}
It gives a set of $s$-stage local symmetries (\ref{301}), 
number of them coincides with defect $\left[\breve{s'}\right]$ of $s$-stage system (\ref{601})  
\begin{eqnarray}\label{909}
\begin{array}{l}
\delta_{\breve s}q^A=
\sum^{s-2}_{p=0}\stackrel{(p)}{\epsilon}{}^{\breve s}
R^{(p)}_{\breve s}{}^A, \qquad \qquad \qquad  
R^{(p)}_{\breve s}{}^1=Q^{(s-p)}{}_{\breve s}{}^1.
\end{array}
\end{eqnarray}
Higher-derivative terms have the structure 
$\delta_{\breve s}q^1=$
$\stackrel{(s-2)}{\epsilon}{}^{\breve s}$$\widetilde E_{(s) \breve s}{}^1+ \ldots$, ~  
$\widetilde E_{(s) \breve s}{}^1\equiv$$\widetilde A_{(s) \breve s}{}^{s-1}$
$\widetilde A_{(s-1) s-1}{}^{s-2}$$\ldots\widetilde A_{(2) 2}{}^1$,    
where by construction
$rank\widetilde E_{(s) \breve s}{}^1=\left[\breve s\right]=\max$.

Let us construct these symmetries for $s=2, 3, \ldots , N+1$. The procedure stops on the stage $s=N+1$, 
since the structure matrices $A, B, C, D$ are not defined for $N+2$. Then total number of 
the symmetries which can be constructed by using of our procedure is
$\sum_{s=2}^{N+1} \left[\breve s\right]=\sum_{s=2}^{N+1} \left[s-1\right]-
\sum_{s=2}^{N+1} \left[\overline s\right]-\sum_{s=2}^{N} \left[s\right]= 
\left[v^{\underline{N+1}}\right]$, 
i.e. coincides with the number of Lagrangian multipliers remaining undetermined in the Dirac 
procedure. All the symmetries obtained are independent in the sence that matrix constructed 
from the bloks $R^{(s-2)}_{\breve s}{}^1, ~ s=2, 3, \ldots , N+1$ has maximum rank by construction 
(in adapted base, higher derivative parts of the transformations do not mix the variables 
$\delta_{\breve s}q^{\breve s}=\stackrel{(s-2)}{\epsilon}{}^{\breve s}+ \ldots$).

Using Eqs.(\ref{902}), it is not difficult to write manifest form of lower-stage 
symmetries for general quadratic Lagrangian action, namely 

\noindent
{\bf{Second-stage symmetries}}
\begin{eqnarray}\label{906}
\delta_{\breve 2}q^1=\epsilon^{\breve 2}\widetilde A_{(2) \breve 2}{}^1,
\end{eqnarray}
{\bf{Third-stage symmetries}}
\begin{eqnarray}\label{907}
\delta_{\breve 3}q^1=-\epsilon^{\breve 3}\left(D_{(3) \breve 3}{}^{\overline 2}
\widetilde A_{(2) \overline 2}{}^1+
C_{(3) \breve 3}{}^2\widetilde A_{(2) 2}{}^1\right)+
\dot{\epsilon}{}^{\breve 3}\widetilde A_{(3) \breve 3}{}^2\widetilde A_{(2) 2}{}^1
\end{eqnarray}
{\bf{4-stage symmetries}}
\begin{eqnarray}\label{908}
\begin{array}{l}
\delta_{\breve 4}q^1=-\epsilon^{\breve 4}\left(\left(D_{(4) \breve 4}{}^{\overline 2}-
C_{(4) \breve 4}{}^3B_{(3) 3}{}^{\overline 2}\right)\widetilde A_{(2) \overline 2}{}^1+
C_{(4) \breve 4}{}^2\widetilde A_{(2) 2}{}^1\right)+ \cr
\dot{\epsilon}{}^{\breve 4}\left(\widetilde A_{(4) \breve 4}{}^3B_{(3) 3}{}^{\overline 2}
\widetilde A_{(2) \overline 2}{}^1-
\left(D_{(4) \breve 4}{}^{\overline 3}\widetilde A_{(3) \overline 3}{}^2+
C_{(4) \breve 4}{}^3\widetilde A_{(3) 3}{}^2\right)\widetilde A_{(2) 2}{}^1\right)+
\ddot{\epsilon}{}^{\breve 4}\widetilde A_{(4) \breve 4}{}^3\widetilde A_{(3) \breve 3}{}^2
\widetilde A_{(2) 2}{}^1.
\end{array}
\end{eqnarray}
Thus the structure matrices $A, B, C, D$ of the Dirac procedure determine independent 
local symmetries of general quadratic Lagrangian action. Number of the symmetries coincides with number of 
Lagrangian multipliers remaining arbitrary in the end of Dirac procedure. 
Surprising conclusion following from the present analysis is that search for gauge 
symmetries of quadratic theory do not requires separation of the Dirac constraints 
on first and second class subsets. We have used rank properties of the matrices 
$\{\Phi_1, \Phi_p\}$ only, whereas structure of the brackets $\{\Phi_k, \Phi_p\}, ~ k, p > 1$ turns out to 
be irrelevant for the process.

\section{Acknowledgments}
Author would like to thank the Brazilian foundations CNPq and FAPEMIG 
for financial support.

\end{document}